# A Model for Android and iOS Applications Risk Calculation: CVSS Analysis and Enhancement Using Case-Control Studies

Milda Petraityte1, Ali Dehghantanha2, Gregory Epiphaniou3

1,2: School of Computing, Science and Engineering, University of Salford- United Kingdom
3: School of Computer Science and Technology, University of Bedfordshire – United Kingdom

milda.petraityte@hotmail.com, a.dehghantanha@salford.ac.uk, gregory.epiphaniou@beds.ac.uk

**Abstract.** Various researchers have shown that the Common Vulnerability Scoring System (CVSS) has many drawbacks and may not provide a precise view of the risks related to software vulnerabilities. However, many threat intelligence platforms and industry-wide standards are relying on CVSS score to evaluate cyber security compliance. This paper suggests several improvements to the calculation of Impact and Exploitability sub-scores within the CVSS, improve its accuracy and help threat intelligence analysts to focus on the key risks associated with their assets. We will apply our suggested improvements against risks associated with several Android and iOS applications and discuss achieved improvements and advantages of our modelling, such as the importance and the impact of time on the overall CVSS score calculation.

## 1. Introduction

The problematics of IT risk management often circle vulnerabilities within IT systems and behaviour of people who use those systems [1]. It is, therefore, hard to disagree that "there is no 100% security, there is only risk management"[2]. However, one can only make informed decisions about risks as effectively as informative and accurate are the metrics used for risk evaluation.

Common Vulnerability Scoring System (CVSS) is used to calculate the severity of vulnerabilities [3] and risks related to IT assets [4]. Industries rely on CVSS as a standard way to capture the principal characteristics of vulnerabilities and produce a numerical score reflecting their severity. CVSS score is currently used to help the organisations to prioritise their vulnerability management process, yet the primary purpose of the CVSS is to detail the technical aspects of the vulnerability [4]. Furthermore, there is an expectation from organisations to use the CVSS calculation if they want to meet the requirements of certifications. For example, Payment Card Industry Data Security Standard (PCI DSS) requires to 'perform quarterly external vulnerability scans using an Approved Scanning Vendor (ASV) approved by the Payment Card Industry Security Standards Council (PCI SSC)' [5]. The ASV requirements clearly state that 'wherever possible, the ASV must use the CVSS base score for the severity level' [6]. The ASV reports need to be configured in such a way that a card vendor could immediately notice non-compliance and fix any issues 'to be considered compliant, a component must not contain any vulnerability that has been assigned a CVSS base score equal to or higher than 4.0.' [6]. The Telecommunication Standardization Sector (ITU-T) base their Recommendations to technical, operating and tariff questions on CVSS too [7], [8]. Previous recommendations have focused solely on CVSS [7], yet the latest report has been dedicated a Common Weakness Scoring System [8], which is largely focused on the methodology of CVSS and adopts a similar calculation [8]. CVSS score is used alongside Common Vulnerabilities and Exposures (CVE) numbers. MITRE considers CVE as the industry standard to systematically register all discovered software vulnerabilities [9]. Before vulnerabilities are registered in a database they are assigned a CVE number and a CVSS score which can be used for software product compatibility testing, within threat intelligence community to build security alerts and within public watch lists such as Open Web Application Security Project (OWASP) to rank and manage risks [9].

However, since the introduction of CVSS there have been numerous complaints and suggestions for improvement of its calculation [3], [4], [10]–[13]. The CVSS v.2 was recently updated, and several changes were introduced, however, several researchers have shown that this scoring system does not represent the real level of associated risks as well. CVSS exploitability metrics were shown to not correlate strongly with the existence of exploits and have a high false positive rate [4], [11]. Also, grouping vulnerabilities into High, Medium and Low does not represent the distribution of the scores well [14]. Finally, the guidelines of CVSS scoring are not being followed [15]. As a result, the data in National Vulnerability Database (NVD) which contains the latest vulnerabilities with their CVSS scores have poor prediction capability [16].

In addition to the listed issues, neither the documentation of CVSS v2 nor CVSS v3 contains a justification of constant values that have been assigned to the components within the CVSS formula [17], [18]. The changes within the vulnerability ecosystem, such as availability of exploits, patches and updates suggest that vulnerabilities do not retain the same risk score over time. While servility of vulnerabilities is changing over time, CVSS sub-scores are not sensitive to the variation of risk over time. The rapid changes in the Information Security (IS) environment require a significant improvement in CVSS score calculation. As stated in ISO27001 standard any method that is used for risk assessment should ensure that its results are consistent, valid and comparable [19]. Finally, CVSS v2 is not tailored to risks imposed to mobile applications [20].

Mobile devices and their security are quickly becoming a widely recognised issue due to various ways that they can be exploited [21]–[23]. A wide range of mobile malware forensics and threat hunting models have been developed to fight against growing security issues in mobile devices [22], [24], [25]. For example, investigation of attackers remnants on mobile social networking applications [26], [27], Voice over IP (VoIP) apps [28], mobile cloud storage apps[29], and techniques for mobile malware detection [30].

Mobile device-related vulnerabilities are also different from those of traditional personal computers and laptops [30]. A particular mobile device comes with a pre-installed operating system (OS) which cannot be changed or replaced. Besides, all applications are available via a dedicated application store i.e. Google or Apple app stores. These applications undergo provenance – a process of application implementation scrutiny for security [31], [32] before they become available for download. However, it was shown that manufacturers do not always provide patches promptly or sometimes do not provide them at all [20], [33].

In this paper, we explain how the formula of CVSS should be adjusted to more precisely reflect risks associated with applications vulnerabilities and we explore our model applicability by practising it on different mobile device vulnerabilities. We discuss the improvements both on impact and exploitability sub-scores, but most importantly we introduce the concept of time which reflects how the risk score changes over time and when there is an environmental change, i.e. when a proof of concept exploit becomes available or when a patch is released.

The structure of this paper is as follow. The first part focuses on the analysis of CVSS score where it is broken down into its components, the Impact and Exploitability sub-scores. This part, also explains the data sets of this research as well as its collection and analysis. In the second part, we discuss the proposed improvements and introduce the new suggested formula and discuss implementation results. The third part covers the conclusions with suggestions for future work and identified limitations of this study.

## 1.1. Background

Both CVSS v2 and v3 consist of three parts: base score, temporal score and environmental score. The base score is the CVSS score assigned once a vulnerability is evaluated. Both CVSS v2 and v3 consider temporal and environmental metrics as optional, and they are not incorporated into the final CVSS score [17], [34]. Furthermore, temporal score still consists of constant values, yet does not have an explanation of the reasoning behind such values [18].

CVSS calculation is modelled according to the risk formula, where the impact is multiplied by likelihood, whereas CVSS calls it the sub-metrics of impact and exploitability [17]:

$$\text{CVSS} = \text{Impact} \times \text{Exploitability} \qquad (1)$$

In formula 1, the impact measures how an exploited vulnerability could affect an asset and exploitability measures the current state of exploit techniques or code availability, moreover, the vulnerability score becomes higher if it is easy to exploit it. Both impact and exploitability scores can be assigned as pre-defined values shown in Table 1.

Table 1. Values of Impact and Exploitability sub-scores of CVSS v2 [17]

| Impact | | Exploitability | |
|---|---|---|---|
| Confidentiality (C) | | Access Vector (AV) | Network (N), Adjacent Network (A), Local (L) |
| Integrity (I) | Complete (C) Partial (P) None (N) | Access Complexity (AC) | High (H), Medium (M), Low (L) |
| Availability (A) | | Authentication (Au) | None (N), Single (M), Multiple (M) |

### 1.2. Impact sub-score

The impact sub-score is evaluated based on the vulnerability effects on the CIA (Confidentiality, Integrity and Availability) triad [17]. For impact sub-score, CVSS v2 used values of Complete (C), Partial (P) or None (N) [17], but CVSS v3 has adopted the values of High (H), Low (L) and None (N) [18]. The impact sub-score of CVSS v2 was obtained using the following formula:

$$\text{Impact} = 10.41 * (1 - (1 - \text{ConfImpact}) * (1 - \text{IntegImpact}) * (1 - \text{AvailImpact})) \quad (2)$$

Even though the naming of these values in CVSS v3 have been changed and some clarification was introduced into the new definitions, it has not uprooted the key issue of the evaluation itself.

The evaluation of 'None' means no impact to any of the CIA components, while 'Complete' shows a total compromise of all CIA triad. The definition of 'None' and 'Complete' suggests that the definition of 'Partial' is very broad. It could be that if there is a compromise of a component, but it is not complete, then it automatically falls into the category of 'Partial'.

Such evaluation introduces a high possibility that the majority of vulnerabilities will contain an assessment of 'Partial' for all or more than one of the components of their impact sub-score. This definition cannot adequately estimate what is the real impact of a vulnerability as the scale of 'Partial' becomes very broad, namely from minimum to near high. The new definition of 'High' in CVSS v3 includes not only the total compromise of a component but also a significant impact to either of them. However, it is still difficult to draw the line and to identify what kind of compromise is a significant one and what could be the one of a lower impact.

The division into the CIA values should make it easier to calculate the Impact score and to introduce more variety, however, this does not seem to work as Table 2 shows a significant amount of greyed-out combinations which are not used within the data set. Table 3 and Table 4 shows a condensed version of Table 2 that omits the

Availability score for a better display of the statistical score and demonstrates the percentage of frequency in which these combinations are currently being used.

**Table 2:** Incidence of impact for Android (left) and iOS (right) NVD datasets, where impact sub-score values demonstrate the combination of possible CIA combinations and their distribution across the data sets. Greyed out areas in the combinations never occurred within the datasets.

| C | I | A | # | Incidence | C | I | A | # | Incidence |
|---|---|---|---|---|---|---|---|---|---|
| C | C | C | 442 | 54% | C | C | C | 187 | 22% |
| C | C | P | 0 | 0% | C | C | P | 0 | 0% |
| C | C | N | 1 | 0.121% | C | C | N | 0 | 0% |
| C | P | C | 0 | 0% | C | P | C | 0 | 0% |
| C | P | P | 0 | 0% | C | P | P | 0 | 0% |
| C | P | N | 0 | 0% | C | P | N | 0 | 0% |
| C | N | C | 3 | 0.363% | C | N | C | 0 | 0% |
| C | N | P | 0 | 0% | C | N | P | 0 | 0% |
| C | N | N | 9 | 1.09% | C | N | N | 3 | 0.355% |
| P | C | C | 0 | 0% | P | C | C | 0 | 0% |
| P | C | P | 0 | 0% | P | C | P | 0 | 0% |
| P | C | N | 0 | 0% | P | C | N | 0 | 0% |
| P | P | C | 0 | 0% | P | P | C | 0 | 0% |
| P | P | P | 62 | 8% | P | P | P | 293 | 35% |
| P | P | N | 58 | 7% | P | P | N | 33 | 4% |
| P | N | C | 4 | 0.484% | P | N | C | 1 | 0.118% |
| P | N | P | 1 | 0.121% | P | N | P | 1 | 0.118% |
| P | N | N | 134 | 16% | P | N | N | 173 | 20% |
| N | C | C | 6 | 0.726% | N | C | C | 3 | 0.355% |
| N | C | P | 0 | 0% | N | C | P | 0 | 0% |
| N | C | N | 0 | 0% | N | C | N | 2 | 0.237% |
| N | P | C | 0 | 0% | N | P | C | 0 | 0% |
| N | P | P | 17 | 2% | N | P | P | 4 | 0.473% |
| N | P | N | 54 | 7% | N | P | N | 102 | 12% |
| N | N | C | 7 | 0.847% | N | N | C | 18 | 2% |
| N | N | P | 28 | 3% | N | N | P | 25 | 3% |
| N | N | N | 0 | 0% | N | N | N | 0 | 0% |

**Table 3.** Combinations of Confidentiality and Integrity per dataset: Android, obtained from NVD and EDB dataset.

| C | I | NVD | EDB |
|---|---|---|---|
| C | C | 53.00% | 45.45% |
| C | P | 0.00% | 0.00% |
| C | N | 1.08% | 9.09% |
| P | C | 0.00% | 0.00% |
| P | P | 17.00% | 22.73% |
| P | N | 15.70% | 13.64% |
| N | C | 0.16% | 0.00% |
| N | P | 9.00% | 9.09% |
| N | N | 4.18% | 0.00% |

**Table 4.** Combinations of Confidentiality and Integrity per dataset: iOS, obtained from NVD and EDB own dataset.

| C | I | NVD | EDB |
|---|---|---|---|
| C | C | 54% | 52% |
| C | P | 0% | 0% |
| C | N | 1.453% | 7% |
| P | C | 0.0% | 0% |
| P | P | 15% | 24% |
| P | N | 18% | 10% |
| N | C | 0.726% | 0% |
| N | P | 9% | 7% |
| N | N | 4% | 0% |

### 1.3. Exploitability sub-score

Exploitability sub-score consists of values detailed in the right column of Table 1 and is calculated according to the following formula [17]:

$$\text{Exploitability} = 20 * \text{AccessComplexity} * \text{Authentication} * \text{AccessVector} \quad (3)$$

**Table 5.** Constant values in CVSS v2 calculation formula [17]

| | |
|---|---|
| Access Complexity (AC) | High: 0.35 |
| | Medium: 0.61 |
| | Low: 0.71 |
| Authentication (AU) | Requires no authentication: 0.704 |
| | Requires single instance of authentication: 0.56 |
| | Requires multiple instances of authentication: 0.45 |
| Access Vector (AV) | Requires local access: 0.395 |
| | Local Network accessible: 0.646 |
| | Network accessible: 1 |
| Confidentiality Impact (C) | None: 0 |
| | Partial: 0.275 |
| | Complete: 0.660 |
| Integrity Impact (I) | None: 0 |
| | Partial: 0.275 |
| | Complete: 0660 |
| Availability Impact (A) | None: 0 |
| | Partial: 0.275 |

Both Impact and Exploitability sub-scores consist of constant values and a coefficient that is used to maintain the score between 1 and 10. CVSS v2 constant values are displayed in Table 5. However, CVSS v2 documentation does not provide any justification of why such numbers or coefficients are used for exploitability sub-score calculation [17].

### 1.4. Research Data Set

Reported Android and iOS software vulnerabilities were collected from following well-known vulnerability databases:

- **National Vulnerability Database (NVD)** [35]: the database contains existing vulnerabilities which are registered and assigned a reference number. 826 vulnerabilities were registered for Android and 845 vulnerabilities were related to iOS as of 23 May 2016.
- **Exploit Database (EDB)** [36]: contains known exploits for existing vulnerabilities which might be listed in NVD. A total of 44 entries found for Android platform, out of which 15 have not been linked to a known CVE hence disregarded. Out of 142 iOS exploits in EDB only 25 entries were associated to a known CVE and used in this research.
- **Vulnerability Lab** [37] is another source that maps the registered vulnerabilities in NVD and EDB and helps to identify additional vulnerabilities.
- **Symantec Connect** [38] database was used to map the vulnerability exploit data with available patches and proof of concept.

The CVSS scores of related vulnerabilities detected from previously mentioned datasets were analysed and compared to understand the distribution of CVSS scores across a scale of 1 to 10 according to the current CVSS calculation which was discussed previously. CVSS scores range from 1 to 10, where 1 is the lowest vulnerability score, and 10 is the highest, or the vulnerability is most critical. Looking at the comparison of this data in Figure 1 and Figure 2 it is quite obvious that there are CVSS scores with a particularly high amount of data, while some scores have a visibly low number of data. Wanting to understand the reason behind such distribution the collected dataset spreadsheets, both for Android and iOS, were then used to analyse the CVSS the sub-scores of exploitability and impact, which is explained further in detail.

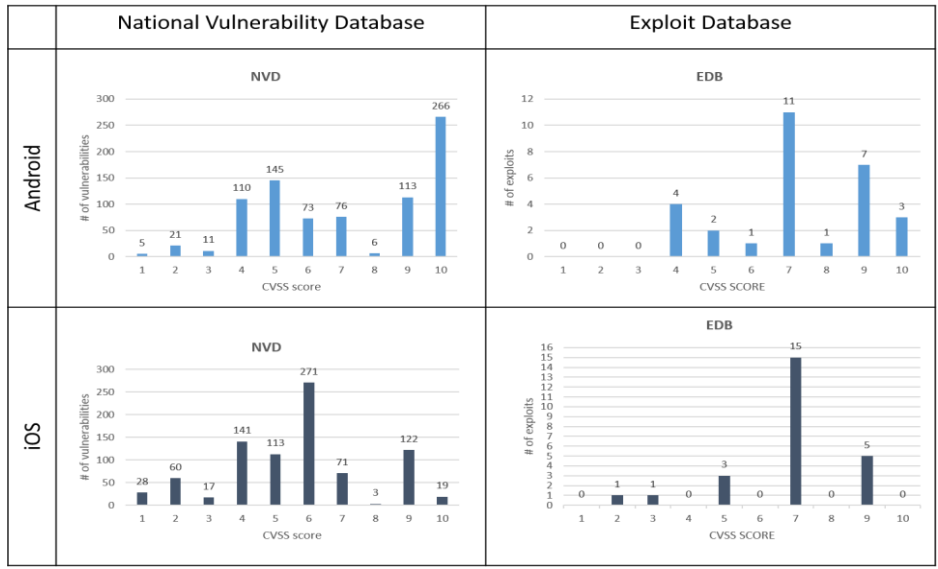

**Figure 1.** CVSS score comparison of Android and iOS vulnerabilities across NVD and EDB

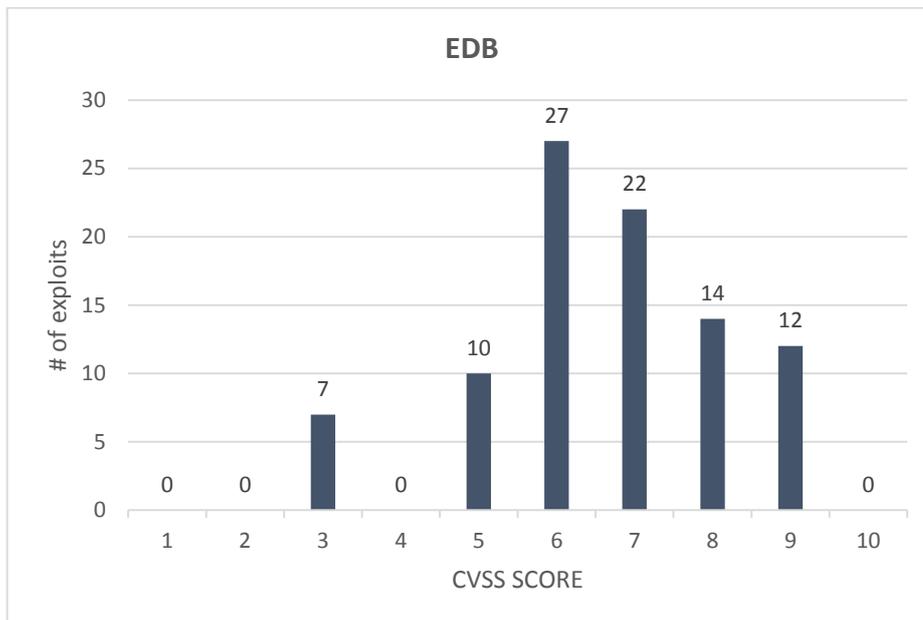

**Figure 2.** Distribution of CVSS scores in Vulnerability Lab

## 1.5. The CVSS Analysis of Data Set

The analysis of the Impact sub-score data displays two problems. First, Figure 1 and Figure 2 display the outcome of a vague evaluation definition. Since the definition of Impact values is not informative enough, it is possible that the features of vulnerability are not evaluated consistently. Therefore the 'Partial' score seems to always fit any average impact to either confidentiality, integrity or availability. Second, since the evaluation of vulnerabilities often does not reflect the amount of corresponding exploits available [4], [11], it is possible that the sub-score values simply do not take into consideration the right features of vulnerability which could be better defined, evaluated and therefore measured.

Figure 3 displays how impact score within our dataset is distributed across the scale of 1-10. It is visible that the score mainly consists of 3 values.

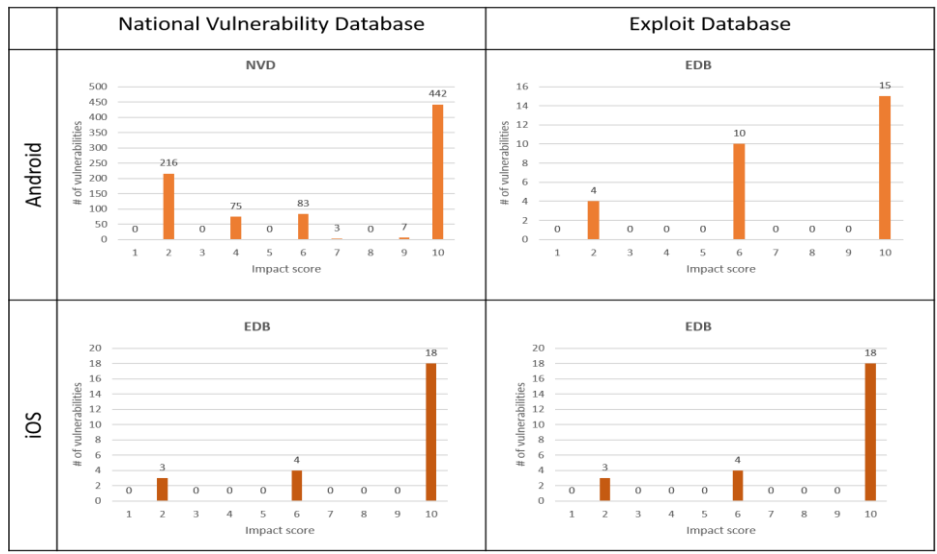

**Figure 3.** Impact sub-score distribution across Android and iOS datasets in NVD and EDB.

The exploitability sub-score within the dataset was analysed to understand its distribution within the scoring scale of 1-10, the way it was done when analysing the Impact sub-score. The results displayed in Figure 4 show that the largest number of vulnerabilities contain an Exploitability score of either 3, or mostly 8 and 10, while other scores are low in number.

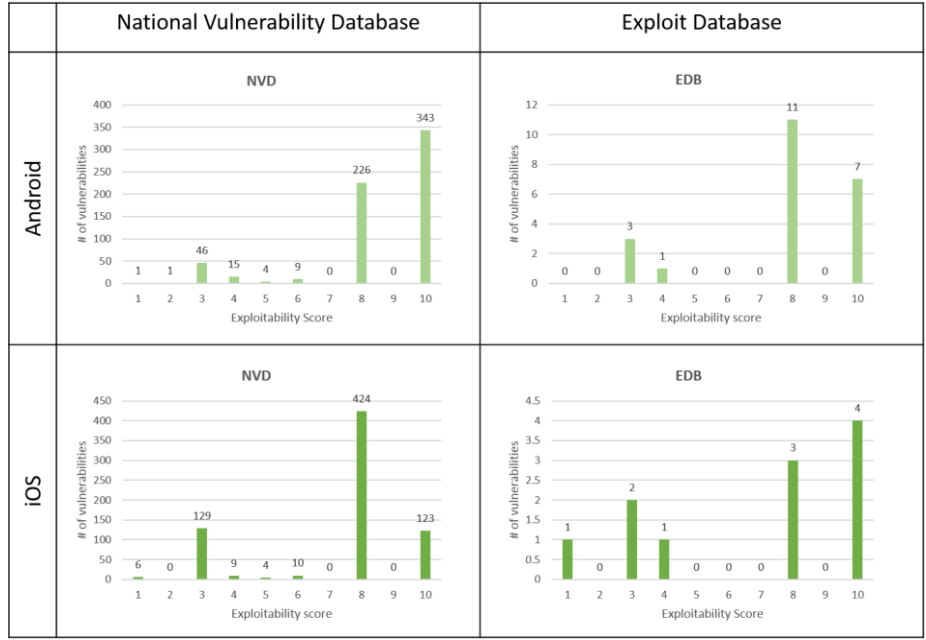

**Figure 4.** Exploitability sub-score distribution across Android and iOS datasets in NVD and EDB.

### 1.6. Proposed Model

In this research we have access to data of existing vulnerabilities and registered exploits, however, this data does not give a drilled down details of a particular use case and how it was remediated if it was at all. It is not possible to know how many devices were affected where a vulnerability was exploited and what happened to it over time. Moreover, the exploit data of mobile devices is very limited, especially for iOS.

The level of variances in this research make it hard to provide comparable results. In such situations, a Randomised Block Design method can be used to compare an individual experiment on several homogenous groups to reduce variances that could impact the study [39]. A comparison is made between two seemingly similar object groups or blocks, where the behaviour of one object group has been analysed and understood. The same features of such two object groups allow to predict and suggest the reaction of the other object group. Therefore, in this research, we compare the data of Android and iOS vulnerabilities and exploits with human viruses and diseases.

There was a variety of studies which successfully adopted the same semantics to describe the disease of a computer system as it was long used to describe a disease in a human body [40]–[42]. In the case of computer systems the system patches and updates is a reaction to any new or existing vulnerability. The concept of changing virus life cycle over time could potentially help to understand how the introduction of time sub-metric makes the score of a vulnerability more accurate [43]–[45]. Overall, the vulnerability lifecycle may have several critical points that could be approximately matched to a biological virus as shown in Table 6.

**Table 6.** Comparison of critical points between a biological virus and a vulnerability

| Biological virus | Vulnerability |
| --- | --- |
| Virus discovery | Vulnerability found |
| Virus is researched/understood | Proof of concept provided |
| Infected individuals reported | Exploit is available |
| Virus treatment discovered | Software patch available |
| Virus treated | Software update pushed to the mobile devices |

Despite the similarities, computer and viruses differ in their survival behaviour. Surviving probability of computer viruses drop sharply over the first two months of their life. Statistically just a small percentage of them cause an outbreak in the computer community, while those viruses that survive still decay over time depending on their type [46]. As for human viruses, some of them (i.e. polio, measles, rubella, etc.) never completely disappear but are dormant due to existing preventive controls (i.e. good hygiene or vaccination), yet when these controls are removed due to any circumstances the viruses infect populations. This behaviour could not be associated with computer viruses since a patch is supposed to provide appropriate protection for a vulnerability and for this reason it should not be possible to infect the system using the same vulnerability.

For this reason, a more suitable function to mathematically reflect the declining spread of a computer virus would be a mathematical function rather than a line, i.e. constant number. Actuarial science has long used Panjer's Recursion Formula (PRF) to calculate cumulative risks for the insurance purposes [47], [48]. In the insurance industry, collective risk models are used to evaluate processes that produce claims over time [47]. In our case, it resembles to risk score over time where variables are independent and identically distributed events that could impact the change of the risk score [47] such as the development of an exploit or the opposite – a system patch. These events are independent in the sense that they are not necessarily generated by the same subject, one event could occur before another, and one could occur without the existence of the other. One of the techniques to calculate these independent events is using PRF as shown in Figure 5.

PRF as shown in formula 4 is a sum of the independent variables that re-occur constantly, where N is a random variable with a value of i. In our case N is some exploits for a particular vulnerability and is independent X, that is time over which an exploit or a number of exploits for that particular vulnerability is available.

$$f(x) = \sum_{i=1}^{N} X_i \qquad (4)$$

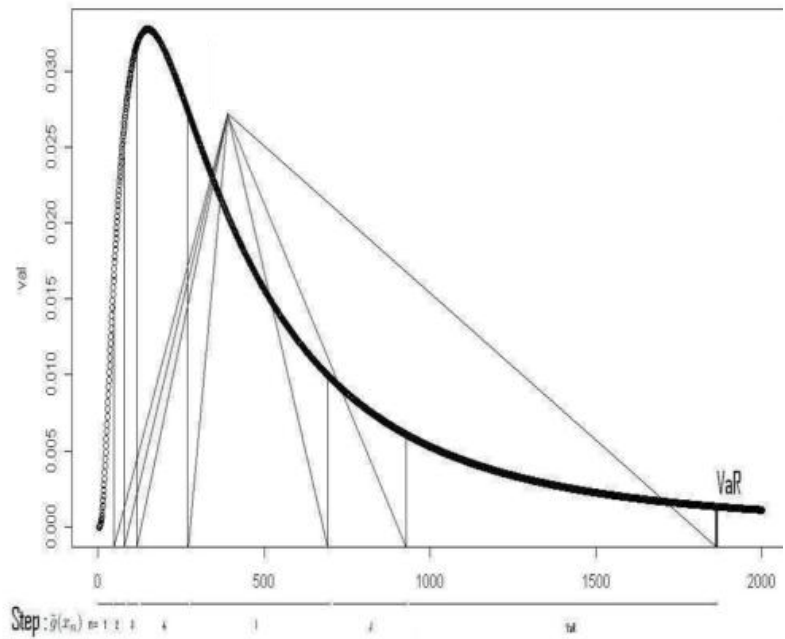

**Figure 5.** The graph of Panjer's Recursion Formula, taken from [48]

Using PRF, we assume that every month there is an equal probability for a new critical point to appear, which is in our case an existence of an exploit or a development of a patch. We also make an assumption that if a new critical point appears in the environment of a vulnerability, it does not mean that there will be no more critical points appearing for the same vulnerability in that month. In other words, the appearance of one critical point is independent from the appearance of any other critical point.

Formula 4 could also be an alternative to Monte Carlo simulation [48]. Its key advantage is that the complete aggregate distribution of claims for a given block of policies may be quickly calculated and the simulation time decreases nearly 60% [48], [49].

PRF is effective when calculating the recursive events, but it has received criticism in the past that the initialisation of the formula is not clear and could introduce variances. Therefore, Guegan & Hassani [48] have modified the algorithm which mixes Monte Carlo simulation, PRF and kernel smoothing by way of introducing either Binomial or Poisson distributions to the core formula. Poisson distribution is suitable to calculate the total amount of events during a particular time span, while Binomial distribution calculates only positive events during that time. Since we are interested in the amount of total events during the lifetime of vulnerability, therefore the inclusion of Poisson distribution to Panjer's PRF is more suitable to satisfy the requirement of the overall CVSS score calculation (see Formula 5). Where e is the exponential limit, $\lambda$ is a mean or an average of critical points per month and $\kappa$ is the probability of the event:

$$S = \sum_{i=1}^{N} X_i \text{, where } X = e^{-\lambda} \frac{\lambda^{\kappa}}{\kappa!}, \text{ where } \lambda \in R^+ \text{ and } \kappa \in N \qquad (5)$$

The formula should be the sum of all critical point components that are available as data for each vulnerability. Hence, taking into consideration all the components of exploitability sub-score.

**1.7. Results and Discussion**

The analysis of the Impact sub-score showed that the CIA evaluation within the Impact sub-score was too broad and too many vulnerabilities could have fallen under an evaluation of 'Partial' impact as it lacked a clear definition. Also, there is a difference in the severity of a vulnerability depending on whether it is an OS vulnerability or the one which exists within a vulnerable application. Therefore, the CIA score evaluation of 'Partial' could be distinguished into two scores, which separate the vulnerabilities apply to applications and OS. For this research, we named the suggested value for application evaluation 'Partial-Application' and the value related to OS 'Partial-System'. The structure together with suggested calculation for each of these values is shown in Table 7. Following the approach of the old CVSS formula the coefficients of 0.461 for the Partial-Application and 0.515 for the Partial-System value were introduced, where coefficient for the vulnerabilities related to OS takes more weight due to their potentially higher impact.

**Table 7.** Suggested breakdown of Impact sub-score evaluation

| CIA Impact | Current constant values | Suggested constant values |
|---|---|---|
| Confidentiality Impact (C) | None: 0 | None: 0 |
|  | Partial: 0.275 | Partial-Application: 0.461 |
|  |  | Partial-System: 0.515 |
|  | Complete: 0.660 | Complete: 0.660 |
| Integrity Impact (I) | None: 0 | None: 0 |
|  | Partial: 0.275 | Partial-Application: 0.461 |
|  |  | Partial-System: 0.515 |
|  | Complete: 0.660 | Complete: 0.660 |
| Availability Impact (A) | None: 0 | None: 0 |
|  | Partial: 0.275 | Partial-Application: 0.461 |
|  |  | Partial-System: 0.515 |
|  | Complete: 0.660 | Complete: 0.660 |

During the implementation of the suggested method, the vulnerabilities within our datasets were separated into either application or OS vulnerabilities. The recommended coefficients were then applied to these vulnerabilities, and a new CVSS Impact sub-score was calculated. After the implementation, another comparison of vulnerability score distribution within the dataset was performed to evaluate how the suggested method changed the values of CVSS score. The result displayed in Figure 6 shows that only the change of a value within one sub-score of CVSS could make a significant difference to the overall evaluation of vulnerabilities and their distribution across the scores.

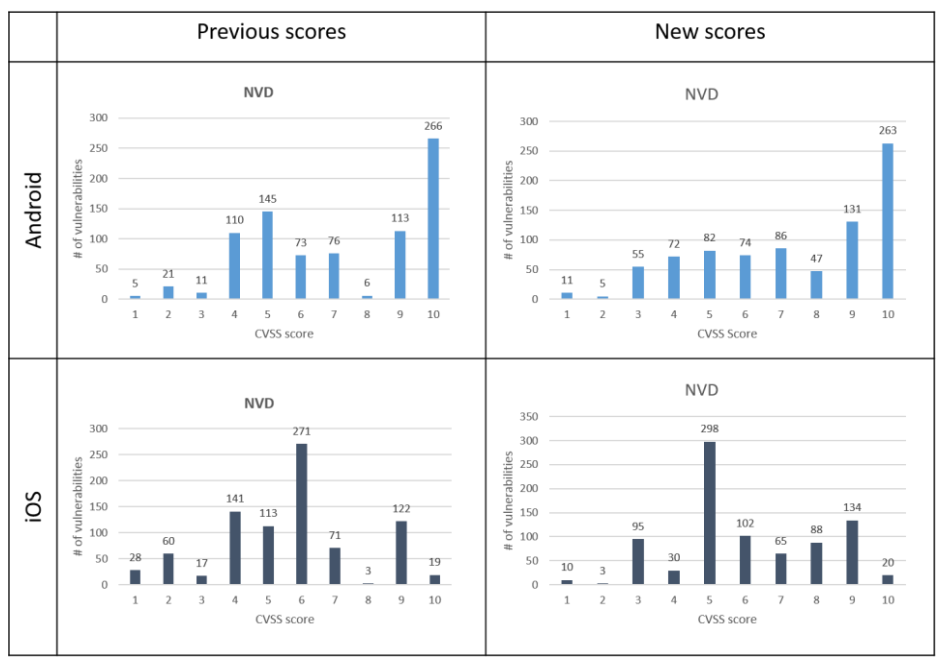

**Figure 6.** Recalculated CVSS scores with an improvement of adding additional value to the Impact sub-score.

However, the split of the 'Partial' score into application or OS vulnerabilities only improved the definition and use of the 'Partial' value of the Impact sub-score. It is possible that the evaluation of 'Complete' should also be reviewed as part of a future research as there are still proportionally high amounts of vulnerabilities with CIA impact of 'Complete' making more than 50% of the entire mobile vulnerability population for both Android and iOS vulnerabilities. This could be one of the reasons why there are large amounts of vulnerabilities with highest CVSS scores.

To improve the Exploitability sub-score, it is necessary to use the PRF Formula 5. The previous discussion suggests that there could be as many as five critical points of vulnerability (see Table 6). However, not all of this information will be available for all vulnerabilities as a vulnerability may not yet have an exploit or a patch. The amount of these critical points would only improve the accuracy of the overall score, without rendering it invalid.

A test with only one critical point would help to back up the above claim. However, to test the effectiveness of the model, we aimed to test the vulnerabilities with as many critical points as possible. Hence, the experiment was applied on vulnerabilities that have an exploit, i.e. are registered with EDB. As of 26 May 2016 Symantec Connect [38] contained three critical points of registered vulnerability exploits within our dataset: availability of proof of concept (POF), an exploit and a patch.

Formula 5 was used to experiment the collected data set run on vulnerabilities that had one critical point, then two critical points and finally three. The Exploitability score was derived according to the age of the vulnerability counting the months since its registration until June 2016. This score was used for calculation of the CVSS base score.

To be able to calculate the new CVSS scores the suggested model was incorporated into the current CVSS base score calculation. The impact sub-score is re-calculated with new CIA values, and PRF was used for calculation of Exploitability as shown in Formula 6.

$$\text{CVSS} = (0.6 * \text{Impact} + 0.4 * \text{Exploitability} - 1.5) * f(\text{Impact}) \qquad (6)$$

Having applied Formula 6 to the data set it is possible to demonstrate that the number of critical points does not make an essential difference to the effectiveness of the Formula 6, yet it provides more accuracy to the overall vulnerability score. Figure 7 displays how a number of critical points improved the accuracy of the calculation and therefore the vulnerability score. However, it does not change the overall principle of how the score changes over several months, as displayed in Figure 7a, 7b and 7c.

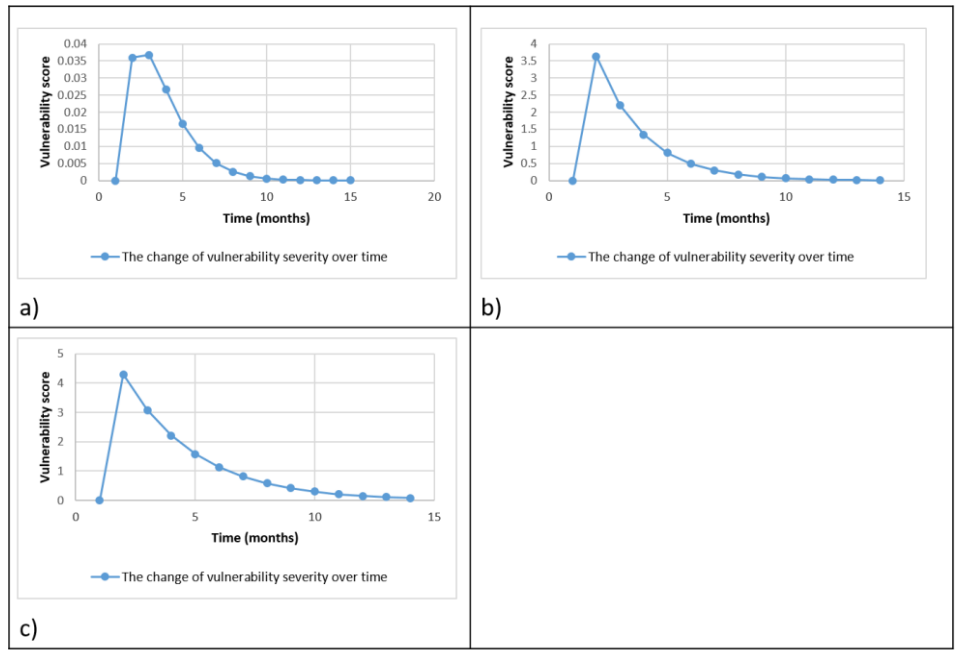

**Figure 7.** CVSS exploitability score change according to the new CVSS calculation model. Graphs display the vulnerability score change over two years (in months) when a) there is one critical point, b) two critical points, c) three critical points discovered

The analysis of the present calculation of the CVSS score has exposed some problems. First, the vulnerabilities do not seem to be measured against their true characteristics which result in vague evaluation system of the overall CVSS score.

Second, the changing technologies require constant monitoring of whether or not the CVSS result of vulnerability is still relevant. Furthermore, a systematic collection of the data showed that there were large amounts of vulnerabilities that retained repeated values across all of their sub-scores, i.e. a CVSS score of 6.8 will always have an Impact sub-score of 6.4 and Exploitability sub-score of 8.6. The analysis of their breakdown values also appeared to be always the same, most likely due to vague vulnerability evaluation system. This could be one of the reasons why the overall distribution of the CVSS scores across the data set was concentrated at only several scores. Therefore the CVSS calculation does not satisfy the purpose that it was designed for.

The suggested Formula 5 introduces a difficulty in updating the scores of existing vulnerabilities as time goes and a critical point is discovered. This would require a strong collaboration between the administration of National Vulnerability Database and Exploit Database team, which is not being done at the moment. It would also require more sophistication in a website design of NVD, where a 'forecast' calculation could potentially be introduced, showing how vulnerability is expected to decrease over time provided a critical point is not discovered.

## 2.  Conclusions and Future Works

Various researchers have shown that the CVSS score calculation has several drawbacks and does not reflect the real situation of the vulnerability risk. However, a number of industry-wide standards use CVSS score to evaluate cyber security compliance, and there is generally a lot of reliance on the scoring system.

Our suggested model improved calculation and distribution of the CVSS base score. However, the formula should be further improved to entirely replace the constant values within the existing formula with those that are dynamic and true representatives of properties of vulnerabilities. Even though the suggested model improved the way that the mobile device vulnerabilities are calculated, there is potentially an overall limitation in how the nature of vulnerability is perceived and therefore evaluated. It was particularly evident in the way vulnerabilities were evaluated for CIA within the Impact sub-score, where some values were never in use for extracted vulnerabilities. Therefore, a potential future work could focus on investigating what the metrics that would allow an accurate evaluation of vulnerabilities according to the meaningful qualities are.

Another limitation of this research is that we have the data of existing vulnerabilities and registered exploits. However this data does not give a drilled down details of a particular exploit case and how it was remediated if it was at all. Therefore, it is not possible to know how many devices were infected with a virus where a vulnerability was exploited and what happened to it over time. Moreover, the scope of the research is limited to the vulnerabilities related to mobile device software and applications only, and therefore further research should be conducted on different types of data to confirm suggested model effectiveness.

# References

[1] A. Shameli-Sendi, R. Aghababaei-Barzegar, and M. Cheriet, "Taxonomy of Information Security Risk Assessment (ISRA)," Comput. Secur., vol. 57, pp. 14–30, 2016.

[2] W. Ahsford, "Sony data breach: 100m reasons to beef up security," 2011. [Online]. Available: http://www.computerweekly.com/news/1280097348/Sony-data-breach-100m-reasons-to-beef-up-security. [Accessed: 14-May-2016].

[3] H. Li, R. Xi, and L. Zhao, "Study on the Distribution of CVSS Environmental Score," pp. 1–4, 2015.

[4] L. Allodi and F. Massacci, "Comparing vulnerability severity and exploits using case-control studies," (Rank B)ACM Trans. Embed. Comput. Syst., vol. 9, no. 4, 2013.

[5] PCI SSC, "Payment Card Industry (PCI) Card Production: Logical Security Requirements," no. May. 2013.

[6] PCI SSC, "Payment Card Industry (PCI) Data Security Standard: Technical and Operational Requirements for Approved Scanning Vendors (ASVs)," October, vol. 21, no. October. 2010.

[7] ITU-T, "Series X: Data Networks, Open System Communications and Security. Common Vulnerability Scoring System," 2011.

[8] ITU-T, "Series X: Data Networks, Open System Communications and Security. Common Weakness Scoring System," 2015.

[9] MITRE.ORG, "CVE," 2016. [Online]. Available: https://cve.mitre.org/index.html. [Accessed: 04-Sep-2015].

[10] L. Gallon, "On the impact of environmental metrics on CVSS scores," Proc. - Soc. 2010 2nd IEEE Int. Conf. Soc. Comput. PASSAT 2010 2nd IEEE Int. Conf. Privacy, Secur. Risk Trust, pp. 987–992, 2010.

[11] A. A. Younis and Y. K. Malaiya, "Comparing and Evaluating CVSS Base Metrics and Microsoft Rating System," no. 1, 2015.

[12] P. Toomey, "CVSS – Vulnerability Scoring Gone Wrong," 2012. [Online]. Available: http://labs.neohapsis.com/2012/04/25/cvss-vulnerability-scoring-gone-wrong/. [Accessed: 03-Jan-2016].

[13] C. Frühwirth and T. Männistö, "Improving CVSS-based vulnerability prioritization and response with context information," 2009 3rd Int. Symp. Empir. Softw. Eng. Meas. ESEM 2009, pp. 535–544, 2009.

[14] P. Toomey, "CVSS – Vulnerability Scoring Gone Wrong | Neohapsis Labs on WordPress.com," 2012. [Online]. Available: https://labs.neohapsis.com/2012/04/25/cvss-vulnerability-scoring-gone-wrong/. [Accessed: 14-May-2016].

[15] C. Eiram and B. Martin, "The CVSSv2 Shortcomings, Faults, and Failures Formulation." pp. 1–13, 2013.

[16] S. Zhang, X. Ou, and D. Caragea, "Predicting Cyber Risks through National Vulnerability Database," Inf. Secur. J. A Glob. Perspect., vol. 24, no. 4–6, pp. 194–206, Nov. 2015.

[17] FIRST, "CVSS v2 Complete Documentation," 2007. [Online]. Available: https://www.first.org/cvss/v2/guide. [Accessed: 16-May-2016].

[18] FIRST, "CVSS v3.0 Preview," 2014.

[19] British Standard Institution (BSI), "ISO/IEC 27001:2013 - Information technology - Security techniques - Information security management systems Requirements," Br. Stand. Online, no. December 2015, 2015.

[20] D. R. Thomas, Beresford, Alastair R., and A. Rice, "Security Metrics for the Android Ecosystem," in Proceedings of the 5th Annual ACM CCS Workshop on Security and Privacy in Smartphones and Mobile Devices, 2015, pp. 87–98.

[21] F. N. Dezfouli, A. Dehghantanha, B. Eterovic-Soric, and K.-K. R. Choo, "Investigating Social Networking applications on smartphones detecting Facebook, Twitter, LinkedIn and Google+ artefacts on Android and iOS platforms," Aust. J. Forensic Sci., 2015.

[22]     M. Damshenas, A. Dehghantanha, K.-K. R. Choo, and R. Mahmud, "M0Droid: An Android Behavioral-Based Malware Detection Model," J. Inf. Priv. Secur., vol. 11, no. 3, Sep. 2015.
[23]     A. Dehghantanha, N. I. Udzir, and R. Mahmod, "Towards Data Centric Mobile Security," IEEE, no. 7th International Conference on Information Assurance and Security (IAS), pp. 62–67, 2011.
[24]     N. Milosevic, A. Dehghantanha, and K.-K. R. Choo, "Machine learning aided Android malware classification, Computers & Electrical Engineering." [Online]. Available: http://dx.doi.org/10.1016/j.compeleceng.2017.02.013.
[25]     M. Petraityte, A. Dehghantanha, and G. Epiphaniou, "Mobile Phone Forensics: An Investigative Framework Based on User Impulsivity and Secure Collaboration Errors," in Contemporary Digital Forensic Investigations Of Cloud And Mobile Applications, 2016, pp. 79–89.
[26]     M. Najwadi Yusoff, A. Dehghantanha, and R. Mahmod, "Forensic Investigation of Social Media and Instant Messaging Services in Firefox OS: Facebook, Twitter, Google+, Telegram, OpenWapp, and Line as Case Studies," in Contemporary Digital Forensic Investigations Of Cloud And Mobile Applications, 2016, pp. 41–62.
[27]     F. Nourozizadeh, A. Dehghantanha, and K.-K. R. Choo, "Investigating Social Networking applications on smartphones detecting Facebook, Twitter, LinkedIn and Google+ artefacts on Android and iOS platforms," Aust. J. Forensics Sci., 2014.
[28]     T. Dargahi, A. Dehghantanha, and M. Conti, "Forensics Analysis of Android Mobile VoIP Apps," in Contemporary Digital Forensic Investigations Of Cloud And Mobile Applications, 2016, pp. 7–20.
[29]     M. Amine Chelihi, A. Elutilo, I. Ahmed, C. Papadopoulos, and A. Dehghantanha, "An Android Cloud Storage Apps Forensic Taxonomy," in Contemporary Digital Forensic Investigations Of Cloud And Mobile Applications, 2016, pp. 285–305.
[30]     K. Shaerpour, A. Dehghantanha, and R. Mahmod, "Trends in Android Malware Detection.," J. Digit. Forensics, Secur. Law, vol. 8, no. 3, pp. 21–40, 2013.
[31]     I. Mohamed and D. Patel, "Android vs iOS Security: A Comparative Study," 2015 12th Int. Conf. Inf. Technol. - New Gener., pp. 725–730, 2015.
[32]     Google, "Android Security: 2015 Year in Review," 2015.
[33]     F. Daryabar, A. Dehghantanha, B. Eterovic-Soricc, and K.-K. R. Choo, "Forensic investigation of OneDrive, Box, GoogleDrive and Dropbox applications on Android and iOS devices," Taylor Fr. Online, no. 0618 (March), pp. 1–28, 2016.
[34]     First, "Common Vulnerability Scoring System (CVSS-SIG)." [Online]. Available: https://www.first.org/cvss. [Accessed: 02-Jan-2016].
[35]     NIST, "National Vulnerability Database," 2016. [Online]. Available: https://nvd.nist.gov/home.cfm. [Accessed: 05-Aug-2015].
[36]     Offensive Security, "Exploit Database," 2016. [Online]. Available: https://www.exploit-db.com/. [Accessed: 10-Aug-2015].
[37]     Vulnerability Lab, "Mobile Vulnerabilities," 2016. .
[38]     Security Focus, "Symantec Connect," 2016.
[39]     W. C. for S. R. Methods, "Randomized Block Designs," 2006. [Online]. Available: https://www.socialresearchmethods.net/kb/expblock.php.
[40]     R. Pastor-Satorras, C. Castellano, P. Van Mieghem, and A. Vespignani, "Epidemic processes in complex networks," Rev. Mod. Phys., vol. 87, no. 3, pp. 1–62, 2015.
[41]     E. Valdano, L. Ferreri, C. Poletto, and V. Colizza, "Analytical computation of the epidemic threshold on temporal networks," arXiv Prepr., vol. 21005, no. 2, p. 19, 2014.
[42]     G. F. De Arruda, E. Cozzo, P. Tiago, F. A. Rodrigues, and Y. Moreno, "Multiple Transitions and Disease Localization in Multilayer Networks," Final Draft. Submitt. Publ., pp. 1–18, 2016.
[43]     G. F. Brooks, J. S. Butel, and S. A. Morse, Medical Microbiology. 2015.

[44]     Scitable, "Host Response to the Dengue Virus," Scitable, 2014. [Online]. Available: http://www.nature.com/scitable/topicpage/host-response-to-the-dengue-virus-22402106. [Accessed: 14-May-2016].
[45]     A. Boianelli, V. K. Nguyen, T. Ebensen, K. Schulze, E. Wilk, N. Sharma, S. Stegemann-Koniszewski, D. Bruder, F. R. Toapanta, C. A. Guzmán, M. Meyer-Hermann, and E. A. Hernandez-Vargas, "Modeling Influenza Virus Infection: A Roadmap for Influenza Research," Viruses, vol. 7, no. 10, pp. 5274–304, Oct. 2015.
[46]     R. Pastor-Satorras and A. Vespignani, "Epidemic spreading in scale-free networks," Phys. Rev. Lett., vol. 86, no. 14, pp. 3200–3203, 2001.
[47]     R. Kaas, M. Goovaerts, J. Dhaene, and M. Denuit, Modern Actuarial Risk Theory, vol. 53. 2008.
[48]     D. Guegan and B. K. Hassani, "A modified Panjer algorithm for operational risk capital calculations," J. Oper. Risk, vol. 4, no. 4, pp. 53–72, 2009.
[49]     L. Spencer and L. Re, "An Overview of the Panjer Method for Deriving the Aggregate Claims Distribution," 2000.